\documentclass[aps,twocolumn,superscriptaddress,prl,floatfix]{revtex4}

\usepackage{amssymb,amsmath,graphicx}
\usepackage{bm}

\begin{document}

\bibliographystyle{apsrev}
\preprint{}

\title{Mechanical Control of Spin States in Spin-1 Molecules \\ and the Underscreened Kondo Effect\footnote{This manuscript has been accepted for publication in {\it Science}. This version has not undergone final editing. Please refer to the complete version of record at http://www.sciencemag.org/. The manuscript may not be reproduced or used in any manner that does not fall within the fair use provisions of the Copyright Act without the prior, written permission of AAAS.}}

\author{J.\ J.\ Parks}
\affiliation{Laboratory of Atomic and Solid State Physics, Cornell University, Ithaca, NY 14853, USA}
\affiliation{Department of Chemistry and Chemical Biology, Cornell University, Ithaca, NY 14853, USA}

\author{A.\ R.\ Champagne}
\altaffiliation{Present address: Department of Physics, Concordia University, Montr\'eal, Qu\'ebec H4B 1R6, Canada}
\affiliation{Laboratory of Atomic and Solid State Physics, Cornell University, Ithaca, NY 14853, USA}

\author{T.\ A.\ Costi}
\affiliation{Institut f\"ur Festk\"orperforschung, Forschungszentrum J\"ulich, 52425 J\"ulich, Germany}

\author{W.\ W.\ Shum}
\affiliation{Department of Chemistry and Chemical Biology, Cornell University, Ithaca, NY 14853, USA}

\author{A.\ N.\ Pasupathy}
\altaffiliation{Present address: Department of Physics, Columbia University, New York, NY 10027, USA}
\affiliation{Laboratory of Atomic and Solid State Physics, Cornell University, Ithaca, NY 14853, USA}

\author{E.\ Neuscamman}
\affiliation{Department of Chemistry and Chemical Biology, Cornell University, Ithaca, NY 14853, USA}

\author{\\S.\ Flores-Torres}
\affiliation{Department of Chemistry and Chemical Biology, Cornell University, Ithaca, NY 14853, USA}

\author{P.\ S.\ Cornaglia}
\affiliation{Centro At\'omico Bariloche and Instituto Balseiro, CNEA, CONICET, 8400 Bariloche, Argentina}

\author{A.\ A.\ Aligia}
\affiliation{Centro At\'omico Bariloche and Instituto Balseiro, CNEA, CONICET, 8400 Bariloche, Argentina}

\author{C.\ A.\ Balseiro}
\affiliation{Centro At\'omico Bariloche and Instituto Balseiro, CNEA, CONICET, 8400 Bariloche, Argentina}

\author{G.\ K.-L.\ Chan}
\affiliation{Department of Chemistry and Chemical Biology, Cornell University, Ithaca, NY 14853, USA}

\author{H.\ D.\ Abru\~na}
\affiliation{Department of Chemistry and Chemical Biology, Cornell University, Ithaca, NY 14853, USA}

\author{D.\ C.\ Ralph}
\affiliation{Laboratory of Atomic and Solid State Physics, Cornell University, Ithaca, NY 14853, USA}
\affiliation{The Kavli Institute at Cornell, Cornell University, Ithaca, NY 14853, USA}

\date{\today}

\begin{abstract}
\begin{center}
\rule{0.8\textwidth}{.1pt}\\
\end{center}
\noindent{\it \textsf{Controlled stretching of individual transition-metal complexes enables direct manipulation of the molecule's spin states and tests of predictions for the underscreened S = 1 Kondo effect.}}\\

The ability to make electrical contact to single molecules creates opportunities to examine fundamental processes governing electron flow on the smallest possible length scales. We report experiments in which we controllably stretch individual cobalt complexes having spin $S$ = 1, while simultaneously measuring current flow through the molecule. The molecule's spin states and magnetic anisotropy were manipulated in the absence of a magnetic field by modification of the molecular symmetry. This control enabled quantitative studies of the underscreened Kondo effect, in which conduction electrons only partially compensate the molecular spin. Our findings demonstrate a mechanism of spin control in single-molecule devices and establish that they can serve as model systems for making precision tests of correlated-electron theories.
\end{abstract}

\maketitle

The electronic states of atoms and molecules depend on the symmetry imposed by their local environment. A simple manifestation occurs in the attachment of ligands to a transition-metal ion to form a coordination complex, which breaks spherical symmetry and causes splittings within the ion's initially degenerate $d$-orbitals. For a complex having spin $S$~$\geq$~1, additional distortions of the ligands combined with spin-orbit coupling can cause splittings within the initially (2$S$~+~1)-degenerate spin states, giving rise to magnetic anisotropy \cite{1,2}. To study the effects of symmetry-breaking distortions, we stretched individual $S$~=~1 molecules within mechanically controllable devices \cite{3}.  Simultaneous electron transport measurements showed that stretching lifts the degeneracy of the $S$ = 1 ground state and enables control of the magnetic anisotropy. The same devices also enabled tests of predictions \cite{4} for the temperature dependence of the underscreened $S$ = 1 Kondo effect \cite{5,6,7}. 

\begin{figure}[b]
	\label{Figure1}
\includegraphics[width=8.6cm]{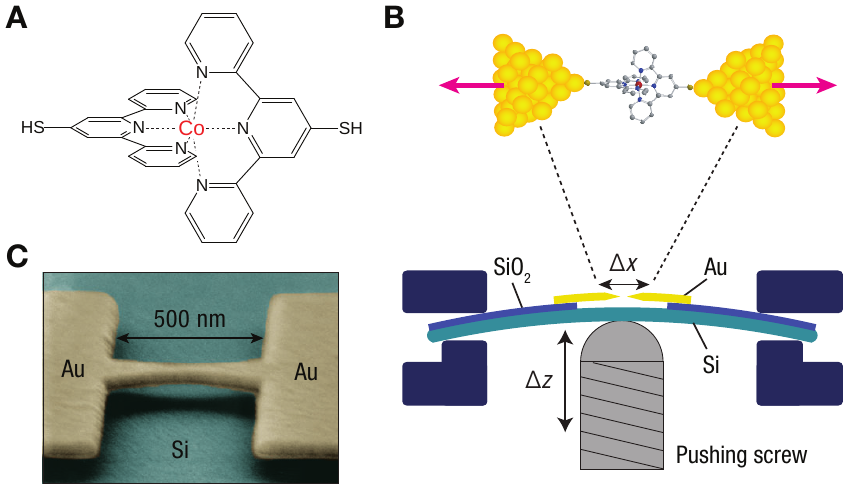} \caption{ \footnotesize{Single-molecule electrical devices based on a transition-metal complex. {\bf (A)} Chemical structure of Co(tpy-SH)$_2$, where tpy-SH = 4'-mercapto-2,2':6',2''-Terpyridine. {\bf (B)} Schematic of the mechanically controllable break junction. {\bf (C)} Scanning electron micrograph of a suspended break junction in false color, prior to molecule deposition.}}
\end{figure}

We studied the Co(tpy-SH)$_2$ complex (Fig.\ 1A), in which a Co ion resides in an environment of approximately octahedral symmetry, through its coordination to six N atoms on two terpyridine ligands. When attached to gold electrodes and cooled to low temperature, this molecule exhibits the Kondo effect \cite{8}; the spin of the molecule is screened by electron spins in the electrodes, leading to a peak in the electrical conductance at zero bias voltage, $V$ \cite{9}. We used the Kondo effect as a spectroscopic probe to interrogate the molecular spin \cite{10}. Previous electron-transport studies of individual metal complexes have probed rich behavior including not only Kondo physics \cite{8, 11,12,13}, but also vibrational excitations \cite{13,14,15} and molecular magnetism \cite{16,17}.

We used mechanically controllable break-junction devices (Fig.\ 1B) to stretch individual molecules while measuring their conductance. The devices were made by fabricating a Au wire suspended above a thin Si substrate (Fig.\ 1C), depositing molecules, and then using electromigration \cite{18} to create a molecular-scale break in the wires before beginning studies in which we tuned the distance between electrodes by bending the substrate. Details, statistics, and discussion of control experiments are provided in the Supporting Online Material, section S1 \cite{19}. We have also measured the same Co(tpy-SH)$_2$ complex using fixed Au electrodes \cite{8,11,13,15,16,17}.

\begin{figure} [t]
	\label{Figure2}
\includegraphics[width=9.0cm]{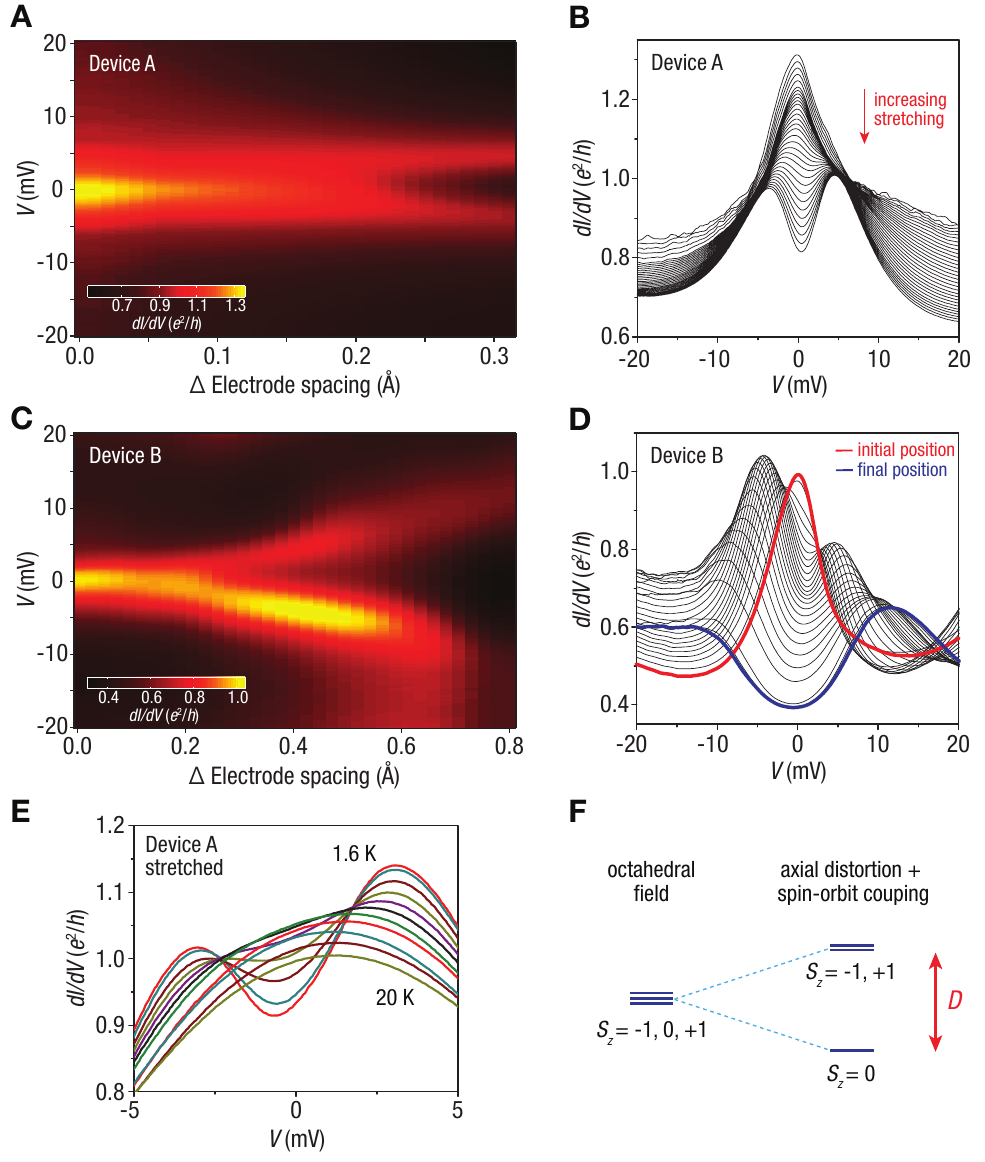} \caption{ \footnotesize{Splitting of the Kondo peaks as a function of mechanical stretching. {\bf (A),(C)} $dI/dV$ as a function of bias voltage and electrode spacing at $T$~=~1.6~K for {\bf (A)} device A and {\bf (C)} device B. {\bf (B),(D)} Line cuts showing $dI/dV$ as a function of $V$ for different values of molecular stretching for {\bf (B)} device A and {\bf (D)} device B. {\bf (E)} Temperature dependence of the conductance for device A on the stretched side of the splitting transition. {\bf (F)} Breaking of the degeneracy among the $S$ = 1 triplet states by uniaxial distortion. }}
\end{figure}

Measurements of differential conductance ($dI/dV$) as a function of increasing electrode spacing at a temperature $T$~=~1.6~K are shown in Fig.\ 2.  At the initial position of the electrodes, the $dI/dV$ spectra exhibited a single peak centered at $V$ = 0 with amplitude of order (but less than) 2$e^2/h$, a signature of Kondo-assisted tunneling through the molecule.  As we stretched each molecule, the single conductance peak split into two beyond a value for the change in electrode spacing that varied from device to device.  For the two devices displayed in Fig.\ 2 and for two others (see section S2 \cite{19}), we could reproducibly cross this transition back and forth between one peak and two.  For the stretched molecules (Fig.\ 2E), the temperature dependence of $dI/dV$ at $V$~=~0 showed a nonmonotonic dependence similar to the $V$-dependence upon increasing $|V|$ from 0.

The observed splitting of the Kondo peak as a function of stretching is in striking contrast to a previous study of the spin-$\frac{1}{2}$ Kondo effect in C$_{60}$ molecules \cite{20}, in which varying the electrode spacing modified the height and width of the Kondo resonance but did not split the peak.  We show that the peak splitting for the Co(tpy-SH)$_2$ complex is caused by a higher-spin $S$ = 1 Kondo effect, together with the breaking of degeneracy within the $S$~=~1 triplet ground state caused by molecular distortion \cite{2}.  For an unstretched $S$~=~1 ion in a ligand field with octahedral symmetry, the triplet states are strictly degenerate according to group theory.  However, if the molecule is stretched axially (the $z$-axis), the $S_z$~=~0 state will be lowered by a zero-field splitting energy $D$ below the $S_z$~=~$\pm$1 states, corresponding to a uniaxial spin anisotropy (Fig.\ 2F).  This broken degeneracy quenches the Kondo resonance near $V$ = 0 and causes conductance peaks at $V$~=~$\pm$$D/e$ due to inelastic tunneling.  The distortion-induced breaking of the ground-state degeneracy for an $S$ = 1 molecule is in contrast to the physics of half-integer spins \cite{21} for which time-reversal symmetry mandates that the ground state remains a degenerate Kramers doublet.  To establish this picture of stretching-induced control of spin states, we show that the spin of our molecules is indeed $S$ = 1, based on temperature and magnetic-field studies, and we verify that stretching produces spin anisotropy by measuring how the level splitting depends on the direction of an applied magnetic field.

The temperature dependence of the Kondo conductance is predicted to depend on the molecular spin $S$ and the number of screening channels in the electrodes.  To be fully screened at zero temperature, a molecule with spin $S$ requires coupling to 2$S$ screening channels \cite{5}.  In our experimental geometry, there are two screening channels consisting of linear combinations of states from the two electrodes with different couplings to the impurity $J_{1}$ and $J_{2}$ that result in two Kondo temperature scales $T_{K1}$ and $T_{K2}$ \cite{22} (we assume $T_{K1}$~$<$~$T_{K2}$).  The Kondo temperatures depend exponentially on the couplings, so in the typical case $T_{K1}$ $<<$ $T_{K2}$.  If $S$~$>$~$\frac{1}{2}$, the result is that over the range $T_{K1}$~$<<$~$T$ $<$ $T_{K2}$ the original spin is only partially screened, to a value $S - \frac{1}{2}$.  Relative to the fully screened $S$~=~$\frac{1}{2}$ Kondo effect, this underscreened effect should produce a much slower rise in the conductance as $T$ decreases below $T_{K2}$ \cite{4,6,7}.  Henceforth we denote $T_{K2}$ as simply $T_{K}$.

\begin{figure}[t]
       \label{Figure3}
       \includegraphics[width=8.6cm]{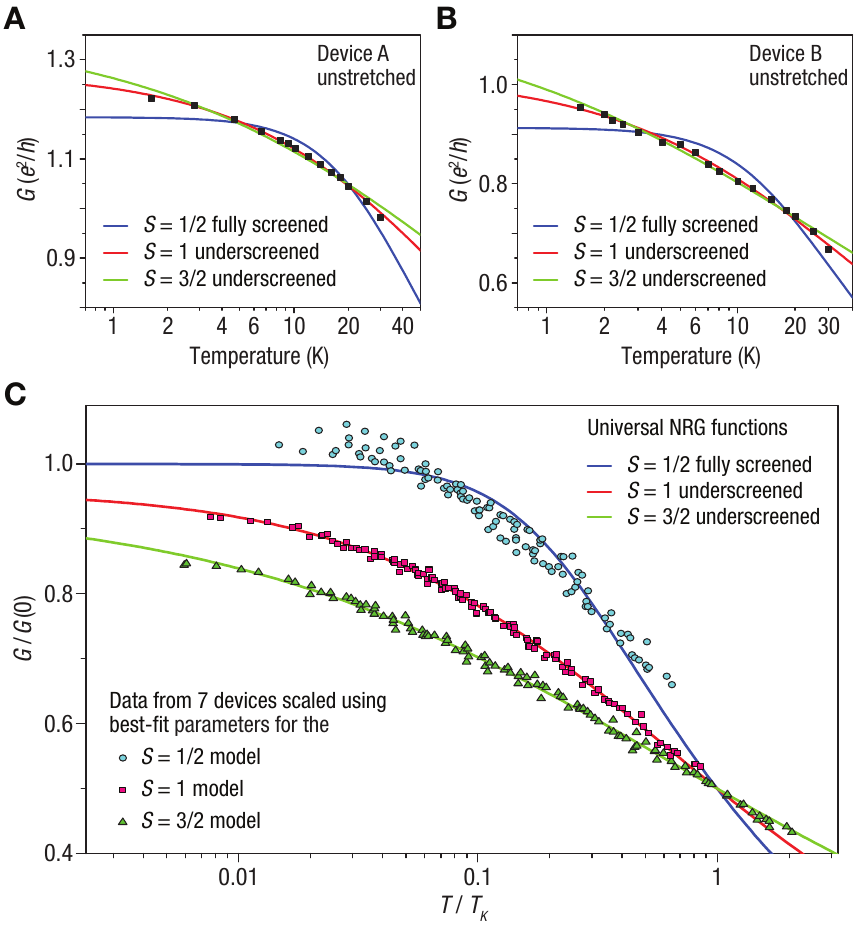}
\caption{ \footnotesize{Temperature dependence corresponding to the underscreened Kondo effect. {\bf (A),(B)} Fits of zero-bias conductance as a function of temperature for {\bf (A)} device A and {\bf (B)} device B to NRG calculations for $S$ = $\frac{1}{2}$, 1, and $\frac{3}{2}$. {\bf (C)} Separate fits to the $S$ = $\frac{1}{2}$, 1, and $\frac{3}{2}$ Kondo models, combining data from seven devices. The collapse of the data on the predicted scaling curve for the underscreened $S$ = 1 Kondo model is excellent; the normalized root-mean-square scatter for the $S$ = $\frac{3}{2}$ ansatz is greater by 44\%, and for $S$ = $\frac{1}{2}$ by 320\%. For the $S$ = 1 fits, the seven Kondo temperatures are 1.1 K, 28 K, 62 K, 81 K, 84 K, 180~K, and 210~K.}}
\end{figure}

Figures 3A and 3B show the measured zero-bias conductance $G(T)$ for the unstretched configuration of devices A and B with fits to numerical renormalization group (NRG) predictions for the fully screened $S$~=~$\frac{1}{2}$ Kondo model \cite{23,24} and the underscreened $S$~=~1 and $\frac{3}{2}$ models \cite{4} (for fitting details, see section S3 \cite{19}).  Each fit has two adjustable parameters, $T_K$ and the zero-temperature conductance $G(0)$.  We find that $G(T)$ deviates strongly from the form for the $S$~=~$\frac{1}{2}$ Kondo effect, and instead agrees quantitatively with the prediction for the $S$ = 1 underscreened case.  We performed the same analysis for 10 unstretched Co(tpy-SH)$_2$ devices (section S4 \cite{19}).  Seven showed similarly unambiguous agreement with underscreened Kondo scaling, and all of these gave superior fits to the $S$ = 1 prediction than to $S$~=~$\frac{3}{2}$.  Discussion of the other 3 samples is provided in section S4.

In Figure 3C, for the 7 devices with underscreened characteristics, we plot $G(T)/G(0)$ versus $T/T_K$ using the parameters $G(0)$ and $T_K$ obtained from separate fits to the $S$~=~$\frac{1}{2}$, 1, and $\frac{3}{2}$ models.  This allows us to test how well the data can all be described by the predicted scaling curves.  We used the root mean square deviation of the data from each theory curve normalized to the average of the scaled data as a goodness-of-fit metric.  The $S$ = 1 fit is best: the deviation of points for the $S$~=~$\frac{1}{2}$ ansatz is greater by 320\%, for $S$~=~$\frac{3}{2}$ by 44\%, and for larger $S$ by more than 80\%.  Additional confirmation that $S$ = 1 is discussed below based on magnetic field studies.  An independent \cite{25} observation of the temperature scaling predicted for the underscreened Kondo effect has also been reported recently for a single $S$ = 1 C$_{60}$ device \cite{26}.

Our interpretation that the stretching-induced Kondo splitting is caused by the breaking of spin degeneracy is confirmed by the presence of spin anisotropy in the stretched state -- the magnetic field ($B$) dependence of the Kondo splitting differs depending on the angle at which $B$ is applied relative to the stretching axis.  Figure 4A shows the expected $B$-dependences of the energy levels in an $S$ = 1 multiplet for $D$~$>$ 0, for several field angles, and Fig.\ 4B displays the energy differences between the first excited state and the ground state, which determine the Kondo peak positions.  For a field orientation approximately perpendicular to the anisotropy axis ($\theta$ = 90$^{\circ}$), when $g\mu_{B}B$~$<<$~$D$ the $B$-dependence of the peak position should be very weak, while when $g\mu_{B}B$ grows large enough to be comparable to $D$, the peak should shift gradually to larger values of $|V|$ with positive curvature.  For $B$ parallel to the anisotropy axis ($\theta$ = 0$^{\circ}$), the $B$-dependence of the peak position should be linear and much stronger.  Our measurements show a dependence on the orientation of $B$ just as expected within this model.  Figures 4C and D show the $B$-dependence for different degrees of stretching for device A, with $\theta$~$\approx$ 90$^{\circ}$.  At a relatively large degree of stretching (Fig.\ 4C), $D$ = 3.45 meV was larger than $g\mu_{B}B$ at our largest field ($\sim$1.5 meV), and the Kondo peak positions were almost independent of $B$.  The slightly negative slope $d|V|/dB$ (corresponding to an effective $g$-factor of 0.21 $\pm$ 0.01) suggests that the stretching axis is not exactly perpendicular to $B$.  For a smaller degree of stretching (Fig.\ 4D), $g\mu_{B}B$ was comparable to $D$ at large fields, and the Kondo peaks shifted weakly to larger $|V|$ as a function of $B$ with positive curvature.  Figures 4E and F show data for $B$ approximately parallel to the molecular axis for fixed-electrode devices, which exhibit zero-field splitting without intentional mechanical stretching (among devices that had Kondo features, $\sim$10\% of the fixed-electrode devices and $\sim$20\% of the adjustable devices exhibited a split Kondo peak at the initial electrode spacing, suggesting that these molecules happened to be strained initially).  In Figures 4E and F, the Kondo peaks shift linearly with $B$, with much larger slopes $d|V|/dB$ than for $\theta$~$\approx$ 90$^{\circ}$, corresponding to $g$-factors of 2.6-2.7 ($\pm$ 0.3).  In device D, the peaks shift initially to smaller $|V|$, pass through zero, and then move to larger $|V|$ (Fig.\ 4F).  From the negative slopes of $d|V|/dB$ in Figures 4C and 4F, we can confirm that $D$ $>$ 0.  The scale of splittings we measure, $\sim$0-5 meV, agrees with our simulations (section S6 \cite{19}) and is typical of zero-field splittings in distorted coordination complexes \cite{2}.

\begin{figure*}
       \label{Figure4}
       \includegraphics[width=12.4cm]{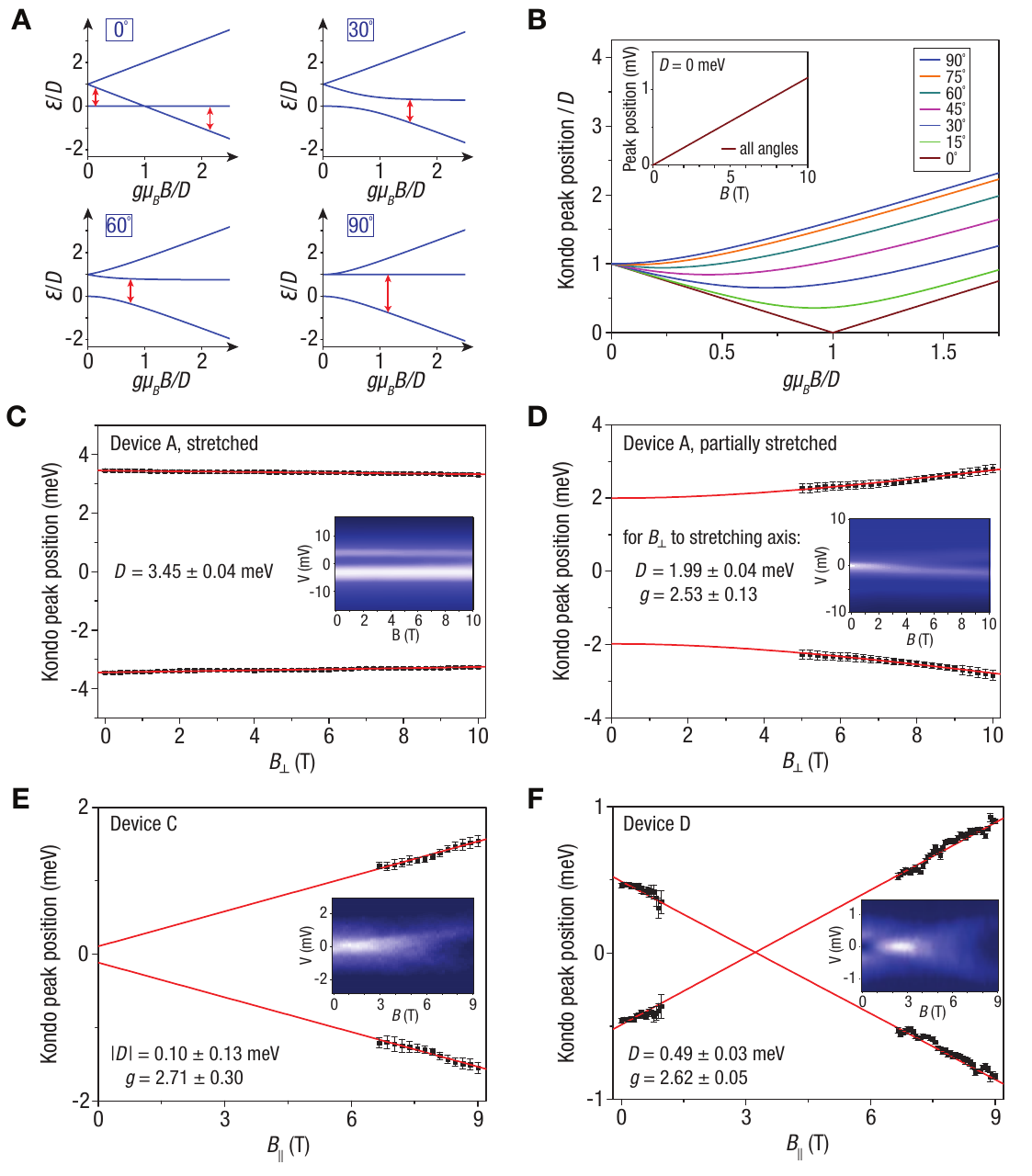}
\caption{ \footnotesize{The magnetic field evolution of the Kondo peaks demonstrates the presence of spin anisotropy. {\bf (A)} Energy eigenvalues (for $S$~=~1) of the model spin-anisotropy Hamiltonian $\mathcal{H} = -g\mu\mathbf{B}\cdot\mathbf{S} + DS_{z}^{2}$ for 4 different field angles with respect to the anisotropy axis. The red arrows indicate the lowest-energy inelastic transitions corresponding to the finite-bias Kondo peaks. {\bf (B)} Predicted Kondo peak position versus magnetic field at several field angles, with energies normalized to the zero-field splitting. {\it Inset:} Kondo peak splitting (for $g$ = 2) as a function of magnetic field for all angles in the absence of spin anisotropy. {\bf (C),(D)} Kondo peak positions versus magnetic field at $T$~=~1.6~K in device A for two different values of stretching, with the magnetic field oriented approximately perpendicular to the molecule axis. The data in {\bf (D)} can be fit for any field angle between 70$^{\circ}$ and 90$^{\circ}$, giving parameters in the range $D$ = 2.0 - 2.2 meV and $g$ = 2.5 - 2.8. {\bf (E),(F)} The magnetic field dependence in fixed-electrode devices C and D with the magnetic field approximately along the molecular axis. The insets contain color plots of $dI/dV$ versus bias voltage and magnetic field. }}
\end{figure*}

These magnetic field data provide further evidence that the spin state of the molecule is $S$ = 1, rather than $\frac{1}{2}$, $\frac{3}{2}$, or any half-integer.  For half-integer spins with $D$ $>$ 0, the ground state for $B$ = 0 is the degenerate Kramers doublet $S_z$~=~$\pm$$\frac{1}{2}$. Regardless of the direction of $B$ or the degree of stretching, this doublet will give rise to a Kondo peak that shifts approximately linearly with $B$ and extrapolates to zero splitting at $B$ = 0 \cite{21}, in contradiction to the data (section S7 \cite{19}).  Integer values of $S$ = 2 can be ruled out based on the temperature scaling described above and because previous studies of six-fold coordinated Co complexes have found almost exclusively $S$~$\leq$~$\frac{3}{2}$, with exceptions only for fluoride ligands \cite{1,2}.  We conclude that only $S$ = 1 can explain the measurements, and from this we identify the charge state of the metal center to be Co$^{1+}$ (section S8 \cite{19}).

In coordination chemistry, the existence of zero-field splittings induced by molecular distortion is well-established, but the ability we demonstrate to continuously distort an individual molecule while simultaneously measuring its zero-field splitting opens the possibility for dramatically more detailed and precise comparisons with theory.  For correlated-electron physics, our results demonstrate that single-molecule electrical devices can provide well-controlled model systems for studying $S$ = 1 underscreened Kondo effects not previously realizable in experiment.  Our work further demonstrates that mechanical control can be a realistic strategy for manipulating molecular spin states, to supplement or replace the use of magnetic fields in proposed applications such as quantum manipulation or information storage \cite{27, 28}.

\end{document}

\onehalfspacing

\begin{center}
{\bf{\Large Supporting Online Material}}
\end{center}

\vspace{0.05in}

\begin{center}
\rule{.85\textwidth}{.25pt}
\end{center}
\begin{itemize}
	\item [] {\bf Contents}
	\begin{itemize}
	{\small\item[S1.] Materials and methods}
	{\small\item[S2.] Additional devices for which the Kondo peak splits with stretching}
	{\small\item[S3.] Fitting to Numerical Renormalization Group (NRG) predictions for $G(T)$}
	{\small\item[S4.] Fitting the measured $G(T)$ to Kondo scaling predictions for different values of $S$, for all samples}
	{\small\item[S5.] Two Co(tpy-SH)$_2$ devices did not show a splitting of the Kondo peak with stretching}
	{\small\item[S6.] Calculations of zero-field splittings}
	{\small\item[S7.] Spin energy levels as a function of magnetic field for $S$~=~3/2}
	{\small\item[S8.] Identification of the charge state of the measured Co(tpy-SH)$_2$}
	{\small\item[S9.] Consideration of alternate mechanisms for the stretching-induced Kondo peak splitting}
	{\small\item[S10.] Calculations of the triplet-singlet energy gap}
\end{itemize}
\end{itemize}
\begin{center}
\rule{.85\textwidth}{.25pt}
\end{center}

\vspace{0.2in}

\noindent{\large{\bf S1. Materials and methods}}\\

To make the mechanically controllable break-junction devices, we start by
fabricating continuous gold lines (32 nm thick, 500 nm long, and with a 50 nm-wide
constriction) suspended 40~nm above a 200 $\mu$m-thick Si wafer \cite{Champagne2005,Parks2007}.
Fabrication of the fixed-electrode devices follows
procedures described in Refs. \cite{Park2002,Jo2006}.  To incorporate the
molecules into either type of device, we clean unbroken wires in an oxygen
plasma to remove organic contaminants and then immerse the samples into a
$<$0.1 mM solution of Co(tpy-SH)$_2$ with PF$_6 ^-$ counterions in acetonitrile, 
allowing the thiol end groups of the molecules to attach to the gold. 
We synthesized the molecules by the process described in Ref. \cite{Park2002} and 
purified them by flash chromatography using an alumina column, 
yielding a final purity $>$99\% prior to deposition.   After the sample
chip is removed from the molecular solution, excess solution is blown 
off with nitrogen gas and the samples
are cooled to low temperature ($T$~=~1.6~K).  We then use electromigration
\cite {Park1999} to create a molecular-scale break in the wires.  The 
electrode motion is calibrated \cite{Agrait2003} from the tunneling conductance of junctions
lacking any added molecules.

After electromigration, we observed a peak in $dI/dV$ at $V$~=~0, characteristic
of the Kondo effect, in approximately 14\% of $\sim$250 mechanically
controllable devices and 18\% of $\sim$200 fixed-electrode devices made with 
Co(tpy-SH)$_2$ solution.  Approximately 3\% of devices exhibited a high-resistance
Coulomb blockade characteristic without a Kondo feature, and the rest had
featureless $dI/dV$ vs.\ $V$ traces indicative of simple tunneling. Among the
devices with a conductance feature near zero-bias, at the initial electrode
spacing $\sim$20\% of the adjustable devices and $\sim$10\% of the fixed-electrode
devices had split peaks (suggesting that the molecule is already
stretched); the peak positions ranged up to $|V|$~=~5~mV.  Four adjustable
devices possessed sufficient stability to repeatedly observe the transition
between split and unsplit peaks as a function of stretching.  No devices showed
a decrease in the Kondo splitting with stretching; two showed a zero-bias Kondo
peak that did not split with stretching (see Section S5). Among 47
adjustable and 24 fixed-electrode devices prepared as control samples with
acetonitrile alone, none contained a zero-bias peak in $dI/dV$ and only one
exhibited Coulomb blockade.

\vspace{0.35in}
\noindent{\large{\bf S2. Additional devices for which the Kondo peak
splits with stretching}}\\

\begin{figure}[h]
	\centering
	\includegraphics[width=0.8\textwidth]{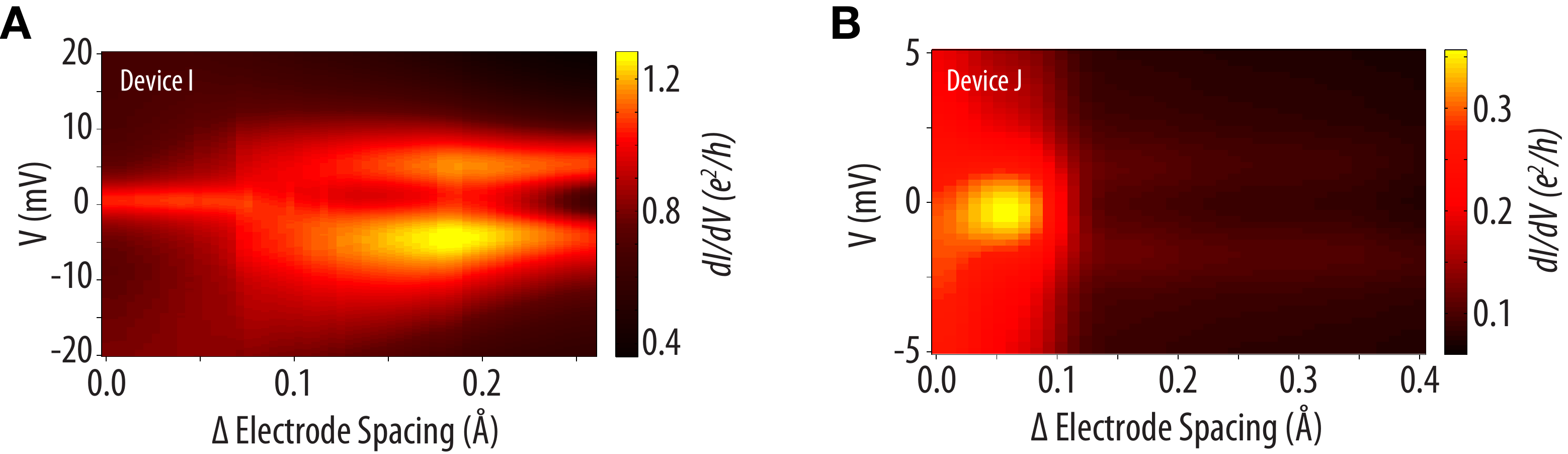}
	\caption{Two additional devices that showed a splitting of
the Kondo peak with stretching.}
	%\vspace{0.6 cm}
       \label{fig:devsplit}
\end{figure}

In Fig.~\ref{fig:devsplit} we show data for two devices (Devices I
and J), in addition to the two described in the main text, that
display a splitting of the Kondo peak with stretching.  While we were
able to tune from a single Kondo peak to split Kondo peaks and then
return to a single peak on a subsequent reverse scan with Devices I
and J, these junctions were not sufficiently stable to make detailed
temperature and magnetic field measurements.

\vspace{0.35in}
\noindent{\large{\bf S3. Fitting to Numerical Renormalization Group
(NRG) predictions for $G(T)$ }}\\

NRG calculations for Kondo-assisted tunneling via an impurity of spin
$S$ coupled to a \emph{single} screening channel predict the
temperature dependence of the conductance in the form of a discrete set of
points,
$G/G(0)$ as a function of $T$ \cite{Mallet2006}.
The calculations were carried out with a logarithmic discretization parameter
$\Lambda=1.5$ \cite{Costi1994} and retaining 1700 states per NRG iteration.
In order to fit experimental data to these NRG
predictions, we first approximate the NRG results by a set of
analytical fitting functions having the scaling form
$G^{*}_S(T) = G(0)\,f_S(T/T_{K})$, where $T_{K}$ is defined so that
$G(T_{K}) \equiv G(0)/2$, as was done in
Refs.~\cite{Costi1994,GoldhaberGordon1998b}.  We do not wish to imply
any physical significance to the form of these fitting functions; they
are merely an attempt to approximate the NRG
results with a minimum number of adjustable parameters.
We use as a starting point the phenomenological Goldhaber-Gordon form
\cite{GoldhaberGordon1998b} for the temperature dependence of the
spin-1/2 Kondo effect
\begin{align}
\label{eqn:g}
g_{S}(T/T_{K}) = \left[1 +
\left(\frac{T}{T'_{K}}\right)^{\xi_{S}}\right]^{-\alpha _{S}},
\end{align}
\begin{figure}[b!]
	\centering
	\includegraphics[width=0.55\textwidth]{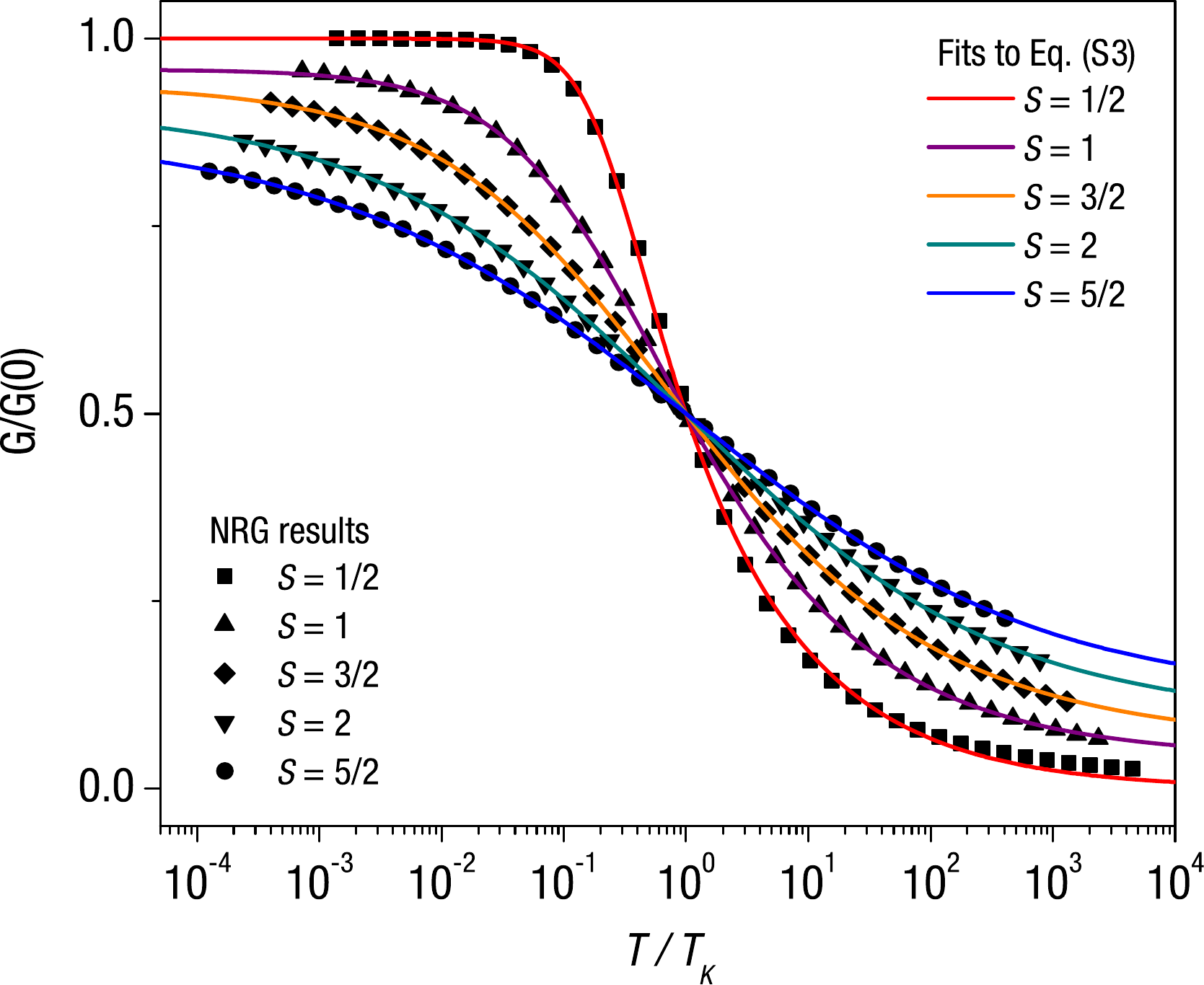}
	\caption{Fits of NRG conductances $G(T)$ to
Eq.~(\ref{eqn:fit}) for $S$~=~1/2, 1, 3/2, 2, and 5/2.}
%	\vspace{0.6 cm}
      \label{fig:NRG}
\end{figure}
but we relax the condition $\xi_{S}$~=~2 (applicable for a fully
screened Kondo effect) since for $S>1/2$, the ground state of an
underscreened Kondo system is not a Fermi liquid.  Note that the
quantity $T'_{K}$ of Eq.~(\ref{eqn:g}) is related to the
actual Kondo temperature by
\begin{align}
T'_{K} = \frac{T_{K}}{(2^{1/\alpha _{S}} - 1)^{1/\xi _{S}}}.
\end{align}
We find that as the temperature is reduced, functions of the form in
Eq.~(\ref{eqn:g}) saturate more quickly to the maximum value of 1 than
do the NRG results for the underscreened Kondo effects -- the
calculated conductance is not yet saturated at the lowest
temperatures $T/T_{K}\sim 10^{-4}$ available in the results.  To
better account for this slow saturation, we find over many decades of
temperature around $T_K$ that good fits to the NRG results can be
obtained by including a small offset parameter $\Delta_{S}$, such that for our
phenomenological scaling functions we use the form
\begin{align}
\label{eqn:fit}
G^{*}_{S}(T) &= G(0) \,\underbrace{\left[(1-\Delta_{S})\, g_{S}(T/T_{K}) +
\Delta_{S}/2\right]}_{f_{S}(T/T_{K})}.
\end{align}
In the above, the coefficient $(1-\Delta_{S})$ helps to account for the
incomplete
saturation of conductance for underscreened Kondo effects at the lowest
temperatures available in the NRG results, and the second term $\Delta_{S}/2$ is
added to retain consistency with the definition of the Kondo temperature;
one can verify by inspection that the term allows the condition
$G^{*}_{S}(T_{K}) = G(0)/2$ to be satisfied.
The reader should not be confused by the fact that $G^*_{S}(0)\neq G(0)$,
because in our fits we will be concerned only about temperatures
within several decades of $T_K$, not with the ultimate
low-temperature limit.

In Fig.~\ref{fig:NRG}, we show fits of Eq. (\ref{eqn:fit}) to the NRG
data for $S$~=~1/2, 1, 3/2, 2, and 5/2, where $\xi _{S}$,
$\alpha_{S}$, and $\Delta_{S}$ are used as fitting parameters.  The
case of $S$~=~1/2 is an exception -- here we fixed $\xi _{S}$~=~2 and
$\Delta_{S}$~=~0 as was done in Ref.~\cite{GoldhaberGordon1998b}.  The fitting
parameters are summarized in Table \ref{table:exponents}.
The NRG results are very well described by
Eq. (\ref{eqn:fit}) for over 7 decades in $T/T_{K}$, with
particularly good agreement in the regime $10^{-3}$~$T_{K} \lesssim T
\lesssim T_{K}$, which is the most relevant for comparison to our
experiments.  In order to fit the predictions of the Kondo models for
different spins $S$ to our experimental data, we employ
Eq.~(\ref{eqn:fit}) keeping fixed the values of the parameters $\xi
_{S}$,  $\alpha_{S}$, and $\Delta_{S}$ as listed in Table
\ref{table:exponents}, and use $G(0)$ and $T_{K}$ as the only
adjustable fitting parameters.  In fitting to the
experimental data, we did not include a background term as an adjustable
parameter in Eq.~(\ref{eqn:fit}).

\vspace{0.5cm}
\begin{table}[h]
\vspace{0.0cm}
\caption{Fitted parameters determined by using Eq.~(\ref{eqn:fit}) to
approximate the NRG predictions for the temperature dependence of conductance
for a Kondo impurity of spin $S$ with a single screening channel
(giving a fully screened Kondo model for $S~=~$1/2 and underscreened for $S$~$>$
~1/2).}
\vspace{0.2cm}
\centering
\begin{tabular}{l c c c}  \hline\hline
\rule{0pt}{3ex}\;\;{\bf Spin} & $\xi_{S}$ & $\alpha_{S}$ & $\Delta
_{S}/2$\\[0.7ex]\hline
\rule{0pt}{3ex}\;\;{\it S} = 1/2\;\;&\;\;2
\;\;&\;\; 0.220$\,\pm\,$0.005 \;\;&\;\;  0 \;\;\\[0.3ex]
		   \;\;{\it S} = 1
\;\;&\;\;0.745$\,\pm\,$0.009\;\;&\;\; 0.506$\,\pm\,$0.009 \;\;&\;\;
0.041$\,\pm\,$0.001\;\;\\[0.3ex]
	         \;\;{\it S} =
3/2\;\;&\;\;0.483$\,\pm\,$0.004\;\;&\;\; 0.670$\,\pm\,$0.008
\;\;&\;\; 0.062$\,\pm\,$0.002\;\;\\[0.3ex]
		   \;\;{\it S} = 2
\;\;&\;\;0.349$\,\pm\,$0.004\;\;&\;\; 0.912$\,\pm\,$0.013 \;\;&\;\;
0.093$\,\pm\,$0.002\;\;\\[0.3ex]
		   \;\;{\it S} =
5/2\;\;&\;\;0.288$\,\pm\,$0.004\;\;&\;\; 1.116$\,\pm\,$0.023
\;\;&\;\; 0.124$\,\pm\,$0.003\;\;\\[1ex]
\hline
\end{tabular}
\label{table:exponents}
%\vspace{0.6 cm}
\end{table}

\vspace{0.35in}
\noindent{\large{\bf S4. Fitting the measured $G(T)$ to Kondo scaling predictions for different values of $S$, for all samples}} \\
\begin{figure}[t!]
	\centering
	\vspace{0.4 cm}
	\includegraphics[width=.99\textwidth]{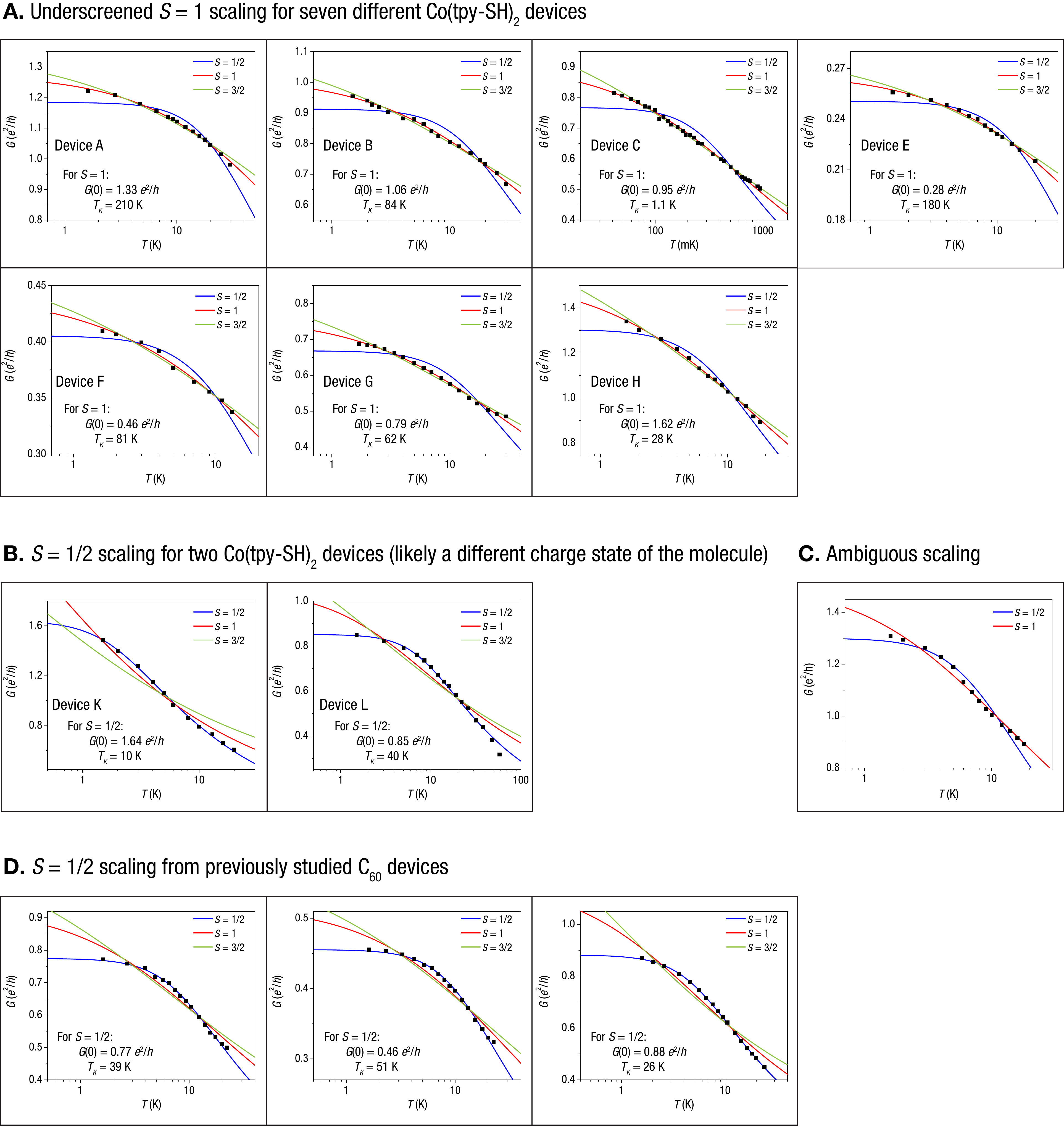}
	\caption{Fits of $G(T)$ to Eq.\ (\ref{eqn:fit}) for ten Co(tpy-SH)$_2$ devices. 
	({\bf A}) Seven devices show a temperature scaling that deviates from the $S$~=~1/2 fully
	screened Kondo prediction, and is instead in good agreement with the underscreened $S$~=~1 Kondo effect.
	({\bf B}) Data from 2 devices are consistent with a spin-1/2 Kondo effect, likely arising from a
	molecule with a Co$^{2+}$ charge state. 
	({\bf C}) One device did not unambiguously fit any of the predicted forms for a single Kondo impurity.
	({\bf D}) $S$~=~1/2 scaling observed in previously studied C$_{60}$ devices \cite{Parks2007}.}
	\vspace{0.3 cm}
	\label{fig:temp}
\end{figure}

We measured the detailed temperature dependence of the unsplit Kondo signal
in ten Co(tpy-SH)$_2$ devices, including two devices with adjustable electrodes (A and B)
and eight with fixed electrodes.  For each set of data $G(T)$ we performed fits
to the predicted scaling curves (Eq.\ (\ref{eqn:fit})) for the fully screened $S$~=~1/2 Kondo model and the underscreened $S$~=~1 and 3/2 Kondo models, with two fitting parameters for each fit, the zero-temperature conductance $G(0)$ and the Kondo temperature $T_K$.  (To be clear, different best-fit values of $G(0)$ and $T_K$ were determined for each value of $S$.)  The results are shown in Fig.~\ref{fig:temp} for all ten unstretched Co(tpy-SH)2 devices, seven of which showed $S$~=~1 scaling, two of which showed $S$~=~1/2 scaling, and one device that showed ambiguous scaling.  For comparison, we also show the same analysis for three previously studied C$_{60}$ devices with $S$~=~1/2 \cite{Parks2007}.

The first conclusion we can draw is that generally there is no ambiguity in distinguishing between realizations of the fully screened $S$~=~1/2  Kondo effect and the underscreened Kondo effects, because the underscreened models predict a \emph{very} slow approach to saturation with decreasing temperature, whereas the fully screened models saturate rapidly at low $T$.  All of the devices we have measured, except for the one exception noted, give excellent fits to either the fully screened $S$~=~1/2 curve or the underscreened curves, with a very poor fit to the other.  From this we conclude, first, that the seven devices shown in Fig.~\ref{fig:temp}A have $S$~$\geq$~1.  That the temperature dependence for each of these samples can be fit well with a single value of $T_K$ indicates that we measure current flow via a single molecule in each device. 

Distinguishing between the different models of the underscreened Kondo effect ({\it e.g.}, $S$~=~1 vs. 3/2) based solely on the temperature dependence is more challenging because the scaling curves are similar, with most of the difference due to a few points at the high and low temperature ends of the fitting ranges.  However, all seven of the samples that exhibit an underscreened characteristic exhibit better agreement with the $S$~=~1 curve than with $S$~=~3/2 or higher spin.  As discussed in the main text, the magnetic field dependence of the Kondo peak splitting provides additional confirmation that $S$~=~1.

For the devices in Fig.~\ref{fig:temp}B that show $S$ = 1/2 scaling, our interpretation is that
the molecule in these devices likely retained a different charge state ({\it e.g.}, Co$^{2+}$) than the molecules exhibiting the underscreened Kondo effect.  The presence or absence of nearby counterions or other charged impurities, or differences in local work functions for the disordered gold electrodes, could easily lead to large differences in the local electrostatic potential at the position of the molecule through which current flows, thereby providing in effect a local gate to change its charge state.  The Co$^{2+}$ state should have a ground state spin $S$~=~1/2 at low temperature \cite{Goodwin2004}, which can produce the conventional fully screened $S$~=~1/2 Kondo effect.

The one device showing a temperature dependence different from both the $S$~=~1 and $S$~=~1/2 predictions (Fig. \ref{fig:temp}C) may possibly be explained by conduction through two molecules in parallel, with different Kondo temperatures or possibly to being in a mixed-valence, rather than a purely Kondo, regime \cite{Schoeller2000}.

\newpage
%\vspace{0.35in}
\noindent{\large{\bf S5. Two Co(tpy-SH)$_2$ devices did not show a
splitting of the Kondo peak with stretching}} \\

\begin{figure}[h]
	\centering
	\includegraphics[width=0.35\textwidth]{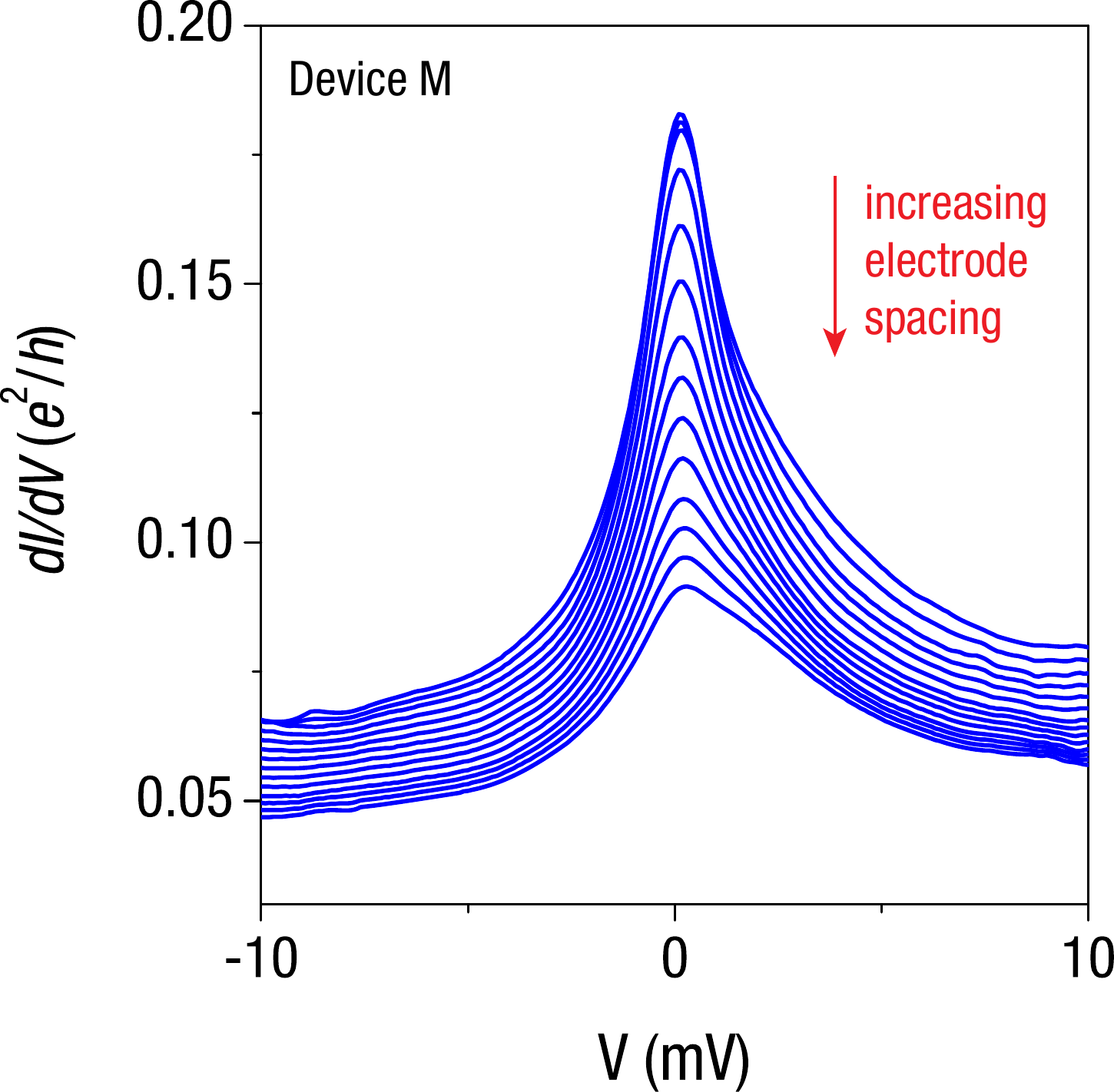}
	\caption{ {\it dI/dV} traces with increasing
electrode spacing for a device at $T$ = 1.6 K, which did not show a
splitting of the Kondo peak with stretching.}
	%\vspace{0.6 cm}
       \label{fig:nosplitKondo}
\end{figure}

Increasing the electrode spacing did not result in a splitting of the
Kondo peak in two devices that we measured. The $dI/dV$ of one such device as a
function of increasing electrode spacing is shown in
Fig.~\ref{fig:nosplitKondo}.  Stretching modifies the height and width of the
Kondo peak, but the peak does not split.  This behavior
might be explained in several ways.  One is that the molecule may not
be strongly bonded to both electrodes so that electrode displacement
does not significantly stretch the metal complex.  The effect of
electrode displacement in this case would be only to reduce the Kondo
conductance and temperature without significantly affecting the symmetry of the molecule
\cite{Parks2007}.  A second possibility is that the spin state of the molecule for these two
devices is $S$~=~1/2 ({\it i.e.}, in the Co$^{2+}$ state), as discussed above.

\vspace{0.35in}
%\newpage
\noindent{\large{\bf S6. Calculations of zero-field splittings}} \\

We carried out calculations of the zero-field splitting for the isolated
molecule spin-triplet state within the DFT coupled-perturbed spin-orbit coupling
formalism \cite{Neese2007}, as implemented in ORCA \cite{ORCA}. Precise reproduction of the
observed experimental $D$ values is challenging because of the dependence on the
molecular environment. It is sensitive in particular to the overall spin density
on the cobalt, which gives some insight into the mechanism of how the variation
in $D$ may be affected at a microscopic level.  For the molecule at equilibrium
geometry, we find that the zero-field splitting is $\sim$0.1 meV, consistent
with the expectation of a small value for nearly octahedral symmetry.  As we tune the
exchange contribution in the density functional to shift the spin density onto
the cobalt atom, the zero-field splitting can increase.  For an exchange
contribution that produces a similar magnetic moment (1.2 Bohr magnetons) on the
cobalt atom as observed in LDA + $U$ calculations on the molecule-electrode
system, we obtain a $D$ of 2.7 meV (using 80\% exchange).  This is of the same
sign and similar in magnitude to the splittings observed in the experiment. The
amount of exchange required is quite high, but the primary effect (to shift
spin density onto the cobalt atom) might arise in the experiment from attachment
to the electrodes.

We also tried to obtain the dependence of $D$ on molecular geometry. This was
complicated by numerical artifacts arising from the solution of the coupled-perturbed
equations treatment. Redistribution of the unpaired electrons between the Co ion and
the ligands might also contribute to changes in the magnitude of $D$ as the 
metal-ligand coupling changes.  However, variations on the order of 1 meV were
calculated. 

\vspace{0.35in}
\noindent{\large{\bf S7. Spin energy levels as a function of magnetic field for $S$~=~3/2}} \\

\begin{figure}[h!]
	\centering
	\includegraphics[width=0.75\textwidth]{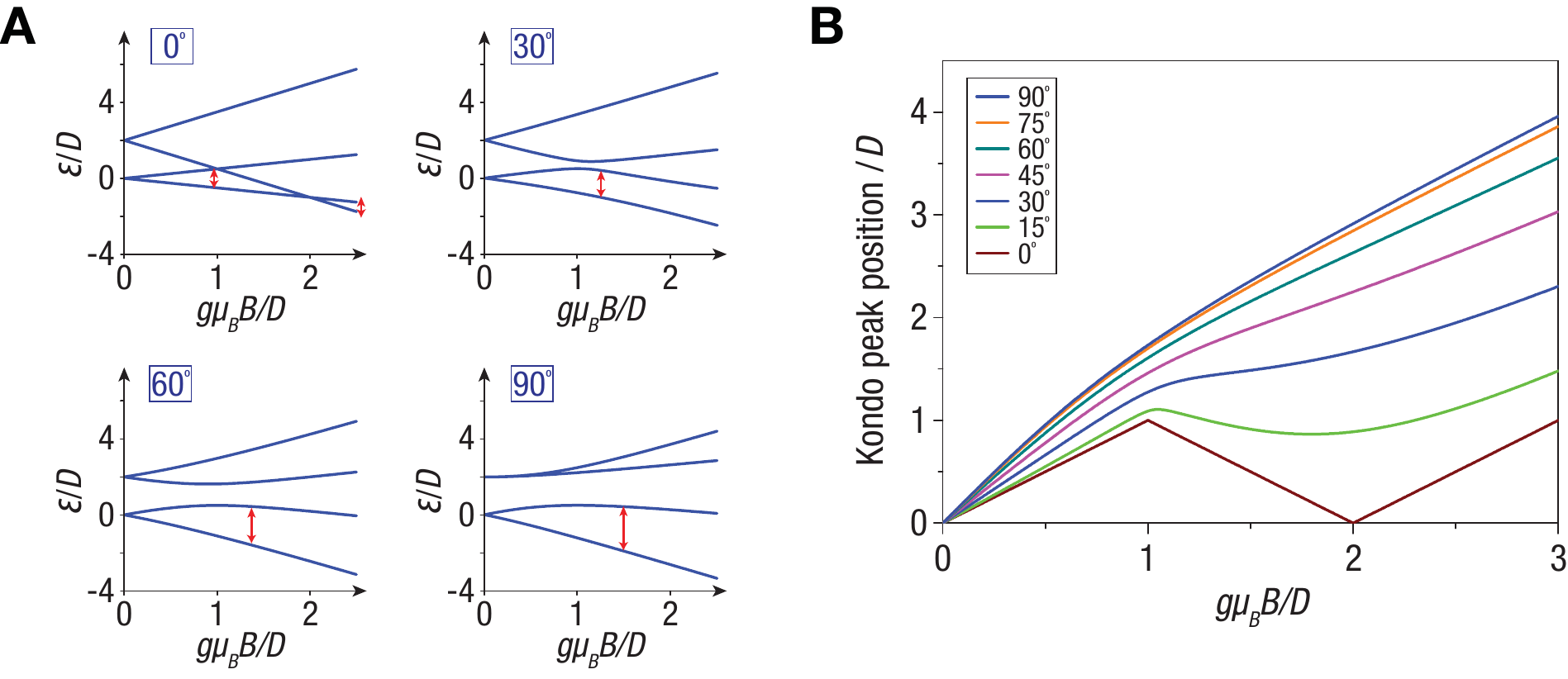}
	\caption{Magnetic field evolution of $S$~=~3/2 energy levels in the presence of anisotropy. ({\bf A}) Energy eigenvalues of the model spin-anisotropy Hamiltonian $H = - g \mu_B \mathbf{B} \cdot \mathbf{S} + D S_z^2$ for 4 different field angles with respect to the anisotropy axis.  The red arrows indicate the lowest-energy inelastic transitions corresponding to the finite-bias Kondo peaks.  ({\bf B}) Predicted Kondo peak position versus magnetic field at several field angles.}
	%\vspace{0.6 cm}
       \label{fig:spin3-2}
\end{figure}

In Figures 4A and 4B of the main manuscript, we plotted the spin energy levels and the Kondo peak position as a function of magnetic field for several different field angles relative to the stretching axis $\hat{z}$, for a simple spin Hamiltonian $H = - g \mu_B \mathbf{B} \cdot \mathbf{S} + D S_z^2$ with $S$~=~1 in the presence of a spin anisotropy energy $D > 0$.  In Fig. \ref{fig:spin3-2} we plot the results of the same model Hamiltonian for $S$~=~3/2.  For any half-integer spin value with $D$~$>$~0, the ground state for $B$~=~0 is a degenerate Kramers doublet with $S_z = \pm 1/2$.  When a magnetic field is applied, no matter what the orientation of $B$ or the magnitude of $D$, the Kramers doublet splits approximately linearly (for $g \mu_B B$~$<$~$D$), and therefore the Kondo peak position would always undergo an increase that extrapolates to zero for $B = 0$.  This is in contradiction to the field dependence that we measure for stretched Co(tpy-SH)$_2$ devices, allowing us to rule out the possibility that $S$~=~3/2, or any other half-integer value.

\vspace{0.35in}
\noindent{\large{\bf S8. Identification of the charge state of the measured Co(tpy-SH)$_2$}} \\

\begin{figure}[h]
	\centering
	\includegraphics[width=0.3\textwidth]{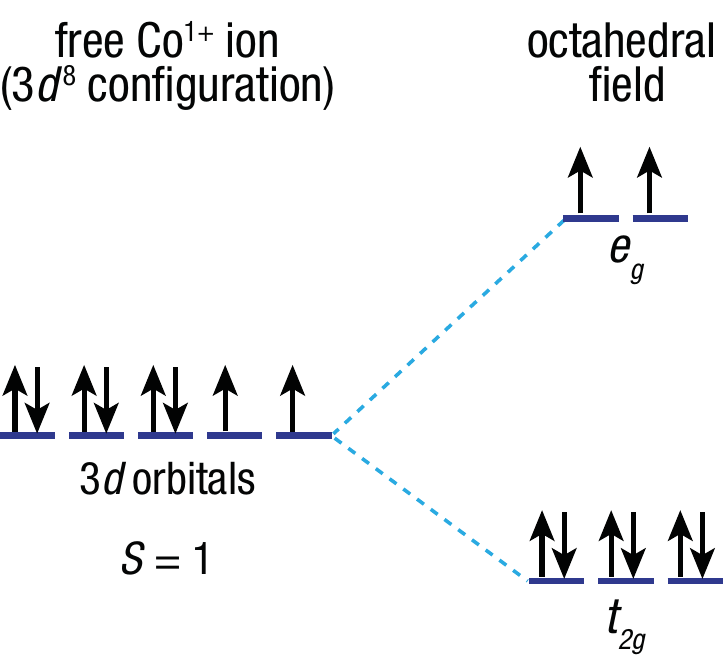}
	\caption{Electronic configuration of the cobalt complex. 
	For a Co$^{1+}$ ion coordinated to ligands with approximately octahedral symmetry, the 5
	metal $d$-orbitals split into a set of $t_{2g}$ and $e_g$ orbitals, and a ground state with $S$~=~1
	is expected.}
%	\vspace{0.6 cm}
	\label{fig:filling}
\end{figure}

We can identify the charge state of the measured Co(tpy-SH)$_2$ complex
based on the determination that the ground-state spin is $S$ = 1.
For a complex in equilibrium with counterions, the metal center would be 
in the Co$^{2+}$ state with an electronic configuration of 3$d^7$ and 
spin $S$~=~1/2 or 3/2 with predominantly $S$ = 1/2 at low temperature
for this complex \cite{Goodwin2004}, inconsistent with the measured 
ground state of $S$~=~1.  Instead, charge transfer of an
electron from the gold electrodes to the complex must result in a 
3$d^8$ Co$^{1+}$ state, which has an $S$~=~1 ground state for an
approximately octahedral ligand field (Fig. \ref{fig:filling}A).  
The other possible low-energy even-electron charge state, Co$^{3+}$,
has $S$~=~0 for 6-fold coordinated complexes \cite{Abragam1986}.

\vspace{0.35in}
\noindent{\large{\bf S9. Consideration of alternate mechanisms for the stretching-induced Kondo peak splitting}} \\

We have considered and can rule out two potential alternative mechanisms for the splitting of the
Kondo peak induced by stretching.  The ``two-stage'' $S$ = 1 Kondo effect \cite{vanderWiel2002,Granger2005}, 
relevant for temperatures less than the Kondo energy scales for both screening channels ($T_{K1}$ and $T_{K2}$),
can lead to a non-monotonic bias and temperature dependence similar to what we observe in the stretched molecules.
However, for this mechanism to explain our data both $T_{K2}$ (reflected in the overall width of the Kondo peak) 
and $T_{K1}$ would have to increase with stretching. This is unphysical because stretching should weaken the coupling
of the metal center to the electrodes.  A splitting of an $S$ = 1 Kondo signal could also result from a 
Jahn-Teller mechanism if distortion caused a non-degenerate $S$ = 0 singlet to shift below the $S$ = 1 triplet to
become the ground state of the molecule.  A triplet-singlet crossing has been suggested as the explanation for 
gate-voltage-induced splittings of Kondo peaks associated with an even number of electrons in GaAs \cite{Kogan2003},
carbon nanotube \cite{Quay2007}, and molecular \cite{Roch2008, Osorio2010} quantum dots.  However, \textit{ab initio} calculations indicate that for the range of stretching in our experiment, the $S$ = 1 triplet state is always
the ground state of Co(tpy-SH)$_{2}$ (see Section S10).  Furthermore, neither a two-stage Kondo effect nor
a triplet-singlet crossing would explain the magnetic-field anisotropies that we measure.

\vspace{0.35in}
\noindent{\large{\bf S10. Calculations of the triplet-singlet energy gap}} \\

\begin{figure}[b!]
	\centering
	\vspace{0.5 cm}
	\includegraphics[width=.85\textwidth]{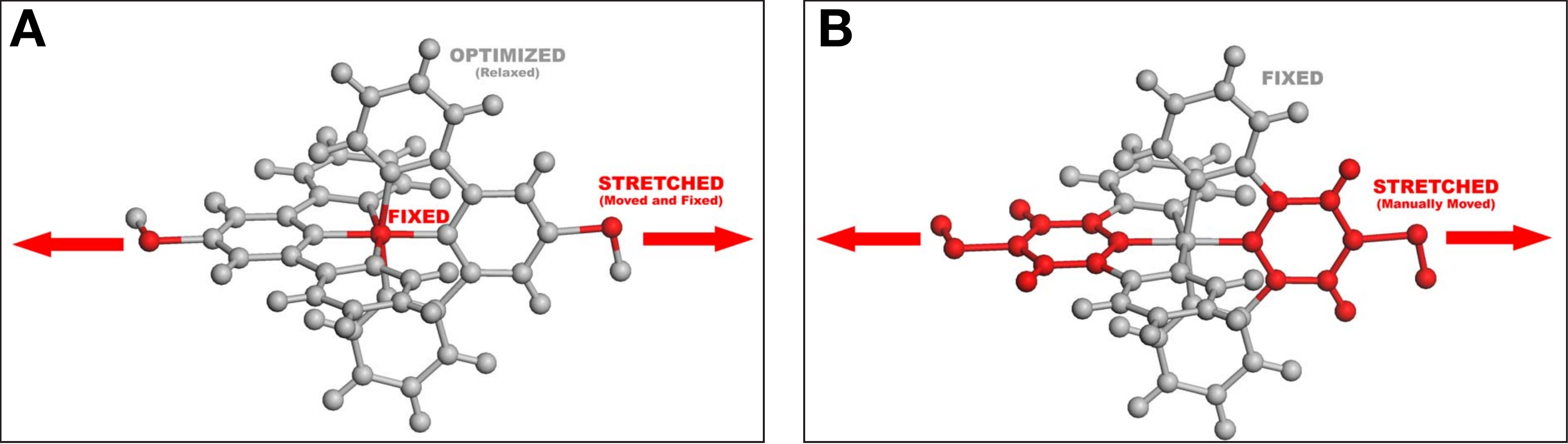}
	\caption{Two simulations of molecule stretching. (\textbf{A}) Stretching
	by increasing the sulfur-sulfur distance while relaxing the positions of
	the other atoms, and (\textbf{B}) stretching by rigidly displacing the
	axial pyridyls.}
	\vspace{0.5 cm}
       \label{fig:moleculesetup}
\end{figure}

We investigated the singlet and triplet energetics of the Co(tpy-SH)$_{2}^{1+}$
complex with and without attachment to electrodes. Isolated molecule
calculations were performed using the ORCA and MOLPRO packages
\cite{ORCA,MOLPRO}; molecule-electrode calculations were performed using the
VASP package \cite{VASP}. We simulated stretching of the molecule in two ways
(i) altering the S-S distance, allowing geometric relaxation of the other atoms,
and (ii) moving the axial pyridyl moieties in a rigid fashion (see Fig.~\ref{fig:moleculesetup}).

\begin{figure}[h]
	\centering
	\vspace{0.3 cm}
	\includegraphics[width=.85\textwidth]{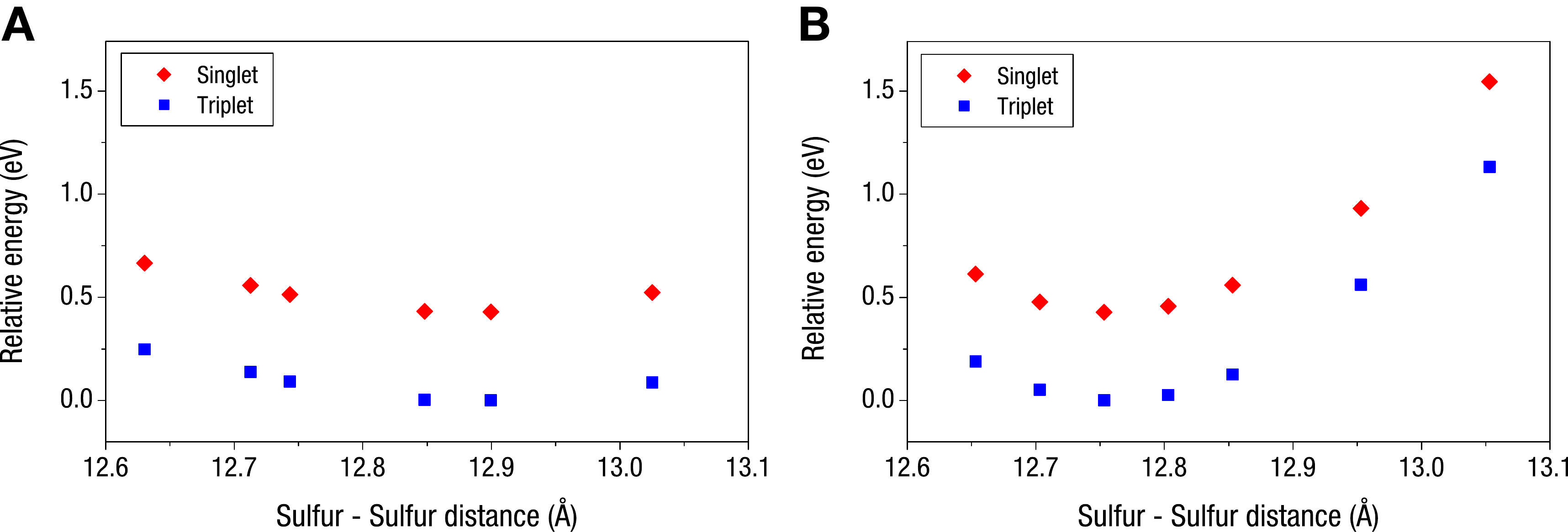}
	\caption{Calculated triplet-singlet energies as a function of stretching
	obtained by (\textbf{A}) varying the sulfur-sulfur distance (with
	relaxation of the other atoms), and (\textbf{B}) rigid displacement of
	the pyridyl groups.  Energies are measured relative to the lowest
	triplet energy.}
	\vspace{0.6 cm}
       \label{fig:S-T}
\end{figure}

Using density functional theory (B3LYP/Def2-TZ2P), we computed the closed-shell
(spin-unpolarized) triplet-singlet gap. This tests the Jahn-Teller type scenario
mentioned in the previous section -- whether mechanical distortion might cause an orbital energy
splitting favoring the singlet over the triplet state. We found, however, that both types
of stretching lead to the triplet state being below the singlet state by about
0.4 eV (Fig.~\ref{fig:S-T}), much larger than any energy scale observed in the
experiment. We also carried out multireference complete active space self-consistent
field (CASSCF/6-31G) calculations using a 14-electron 14-orbital
active space and this yielded a comparable singlet-triplet gap (0.55 eV) near
the equilibrium geometry.

\textit{Electrodes:} To test the effect of attaching to electrodes, we carried
out periodic density functional calculations (PAW, 350 eV plane-wave cutoff)
with the molecule inserted between two model gold electrodes. Here, pure
functionals such as LDA and PBE did not yield magnetization of the molecule in
the triplet state, an artifact of the density functional approximations. Moving
to the LDA + $U$ formalism (Co: $U$ = 8.0 eV, $J$ = 1.0 eV) \cite{Anisimov1991}
recovered a magnetic cobalt atom in the triplet state, with a corresponding
singlet-triplet gap of 0.52 eV at a near-equilibrium molecule-electrode
geometry. This appears consistent with our isolated molecule studies.

\textit{Other low-lying states:} We also investigated the presence of other low-lying 
states in the spectrum. One possibility is an antiferromagnetic (open-shell)
singlet where the cobalt atom is formally Co(III) and the two terpyridine
ligands are formally negatively charged with an unpaired electron,
antiferromagnetically coupled to each other. In the isolated molecule case, we
found the AFM singlet to be lower in energy than the closed-shell singlet. The
AFM singlet-triplet gap is very sensitive to the precise theoretical treatment.
By varying the amount of exact exchange in the density functional, 0\% (BP86),
20\% (B3LYP), 50\% (BHLYP) we observed gaps of 19.8 meV, 3.1 meV, 1.9 meV. In
the case of attached electrodes, we did not find an AFM singlet solution using
the LDA and PBE functionals, likely due to the same self-interaction errors in
the density functionals that led to the disappearance of the magnetism in the
triplet state mentioned above.  Nonetheless, we consider it likely that the AFM
singlet state may be a low-lying state in these systems.